\documentclass[11pt,a4paper]{article}
\usepackage{amsmath,amssymb}
\usepackage{epsfig,graphicx}

\begin{document}

\title{\bf Non-minimal coupling in inflation and inflating with the Higgs
  boson}
\author{F.~L.~Bezrukov\footnote{{\bf e-mail}: fedor@ms2.inr.ac.ru}
\\
\small{\em Max-Planck-Institut f\"ur Kernphysik} \\
\small{\em PO Box 103980, 69029 Heidelberg, Germany}\\
\small{\em Institute for Nuclear Research of Russian Academy of Sciences} \\
\small{\em Prospect 60-letiya Oktyabrya 7a, Moscow 117312, Russia}
}

\date{}
\maketitle

\begin{abstract}
  We analyse the effect of the non-minimal coupling of the form
  $\xi\phi^2R/2$ on the single field inflation.  If the non-minimal
  coupling is large, it relaxes the constraint on the field self
  coupling, making it possible to use the Standard Model Higgs field
  as the inflaton.  At the same time, even small non-minimal coupling
  constant, $\xi\gtrsim10^{-3}$, brings the usual inflaton with
  quartic potential in agreement with the WMAP5 observations.
\end{abstract}

\newcommand{\e}{\mathrm{e}}
\newcommand{\dm}{\partial_\mu}
\newcommand{\GeV}{\,\mathrm{GeV}}
\newcommand{\Mpl}{M_P}

\section{Introduction}
\label{sec:introduction}

It is now widely accepted that the initial phase of the Universe was
that of the exponential expansion, called inflation
\cite{Starobinsky:1979ty,Starobinsky:1980te,Mukhanov:1981xt,Guth:1980zm,%
  Linde:1981mu,Albrecht:1982wi}.  It explains the extreme flatness of
the Universe, and predicts a nearly scale invariant spectrum of the
initial density perturbations, which is now confirmed by experimental
observations of the Cosmic Microwave Background (CMB)
\cite{Komatsu:2008hk}.  Simplest realization of inflation can be made
in a theory of a scalar field.  However, the observed amplitude of the
perturbations (COBE normalization) requires extremely flat potential
for this field, $\lambda\sim10^{-13}$ for the quartic coupling
constant \cite{Linde:1983gd}.  Moreover, simplest models of inflation
(with monomial potentials) predict large amount of tensor modes
generated during inflation.  For the case of quartic interaction this
is already strongly disfavoured by observations.

It was noticed, that non-minimal coupling to gravity relaxes the
required fine-tuning of the coupling constant
\cite{Spokoiny:1984bd,Fakir:1990iu,Salopek:1988qh,Kaiser:1994wj,%
  Kaiser:1994vs,Komatsu:1999mt}.  Here I will argue, that taking into
account non-minimal interaction of the scalar field with gravity
(which is quite natural and is in fact even required by quantum
corrections to the action \cite{Birrell:1982ix}) allows to incorporate
inflation in the Standard Model (SM) without introduction of any new
particles.  Or, if the new scalar inflaton is introduced, non-minimal
coupling largely reduces the amount of tensor modes produced, which
will be bounded by future experiments.

\section{Generic description}

If we write the action for the scalar field and gravity with all
operators of dimension not greater then 4 with no more then two
derivatives\footnote{Terms with more derivatives produce additional
  degrees of freedom in the theory, and presumable need special
  analysis.} we get
\begin{equation}
  \label{eq:1}
  S_{J} = \int d^4x \sqrt{-g} \left\{
    - \frac{M^2+\xi \phi^2}{2}R
    + \frac{\partial_\mu \phi\partial^\mu \phi}{2}
    -V(\phi)
  \right\}
  \;,
\end{equation}
with the potential
\begin{equation}
  V(\phi)=\frac{\lambda}{4}\phi^4+\frac{m^2}{2}\phi^2 \;,
\end{equation}
In the following we suppose that quadratic term is irrelevant during
inflationary regime (it is true if $m\ll M_P$).  The simplest way to
analyse this action (see, eg.\
\cite{Salopek:1988qh,Kaiser:1994vs,Tsujikawa:2004my}) is to make the
conformal transformation
\begin{equation}
  \label{eq:2}
  \hat{g}_{\mu\nu} = \Omega^2 g_{\mu\nu}
  \;,\quad
  \Omega(\phi)^2 = \frac{M^2 + \xi \phi^2}{M_P^2}
  \;,
\end{equation}
where $M_P\equiv 1/\sqrt{8\pi G_N}=2.44\times10^{18}\GeV$ is
the reduced Planck mass.  This transformation leads to a non-minimal
kinetic term for the scalar field, which can be removed by changing to
the new scalar field $\chi$
\begin{equation}
  \label{eq:3}
  \frac{d\chi}{d\phi}=\sqrt{\frac{\Omega^2+6\xi^2\phi^2/M_P^2}{\Omega^4}}
  \;.
\end{equation}
Finally, the action (called the Einstein frame action, opposed to the
original Jordan frame action $S_J$)
\begin{equation}
  \label{eq:4}
    S_E =\int d^4x\sqrt{-\hat{g}} \Bigg\{
    - \frac{M_P^2}{2}\hat{R}
    + \frac{\partial_\mu \chi\partial^\mu \chi}{2}
    - U(\chi)
    \Bigg\}
    \;,
\end{equation}
where $\hat{R}$ is calculated using the metric $\hat{g}_{\mu\nu}$ and
the potential is rescaled with the conformal factor
\begin{equation}
  \label{eq:5}
  U(\chi) =
  \frac{V(\phi(\chi))}{\Omega(\phi(\chi))^4}
  \;.
\end{equation}
Already here one can hope that the situation is better, than without
the non-minimal coupling: for large field values $\Omega\propto\phi$,
and the Einstein frame potential $U$ becomes flat.

In the following two sections I will describe two cases, corresponding
to large and small non-minimal coupling $\xi$.

\section{Large non-minimal coupling, or inflation with the Higgs}

The case of large non-minimal coupling $\xi$ is particularly simple
\cite{Bezrukov:2007ep,Bezrukov:2008cq,Fakir:1990iu,Tsujikawa:2004my}.
We have the following change of variables
\begin{equation}
  \label{eq:chi(h)}
  \chi \simeq \left\{
    \begin{array}{l@{\qquad\text{for}\quad}l}
      \phi
      & \phi<\sqrt{\frac{2}{3}}\frac{M_P}{\xi}
      \;,\\
      \sqrt{\frac{3}{2}}M_P\log \Omega(\phi)
      & \sqrt{\frac{2}{3}}\frac{M_P}{\xi} < \phi
      \;,
    \end{array}
  \right.
\end{equation}
and the potential
\begin{equation}
  \label{eq:U(chi)}
  U(\chi) \simeq \left\{
    \begin{array}{l@{\;\text{for}\;}l}
      \frac{\lambda}{4} \chi^4
      & \chi< \sqrt{\frac{2}{3}}\frac{M_P}{\xi}
      \,, \\
      \frac{\lambda M_P^4}{4\xi^2}
      \left(
        1-\e^{
          -\frac{2\chi}{\sqrt{6}M_P}
        }
      \right)^{2}
      & \sqrt{\frac{2}{3}}\frac{M_P}{\xi} < \chi
      \,.
    \end{array}
  \right.
\end{equation}

Thus, at large $\chi$ the potential is exponentially flat for
\emph{any} value of the self coupling $\lambda$.  The ratio
$\lambda/\xi^2$ defines the energy density at high fields now, and
thus it is possible to satisfy the COBE normalization for any
$\lambda$ by choosing sufficiently large value of $\xi$,
\begin{equation}
  \label{eq:9}
  \xi \simeq \sqrt{\frac{\lambda}{3}}\frac{N_{\mathrm{COBE}}}{0.027^2}
  \simeq  49000\sqrt{\lambda}
  =  49000\frac{m_H}{\sqrt{2}v}
  \;,
\end{equation}
where $N_\mathrm{COBE}\simeq60$.

\begin{figure}
  \centering
  \includegraphics[width=\textwidth]{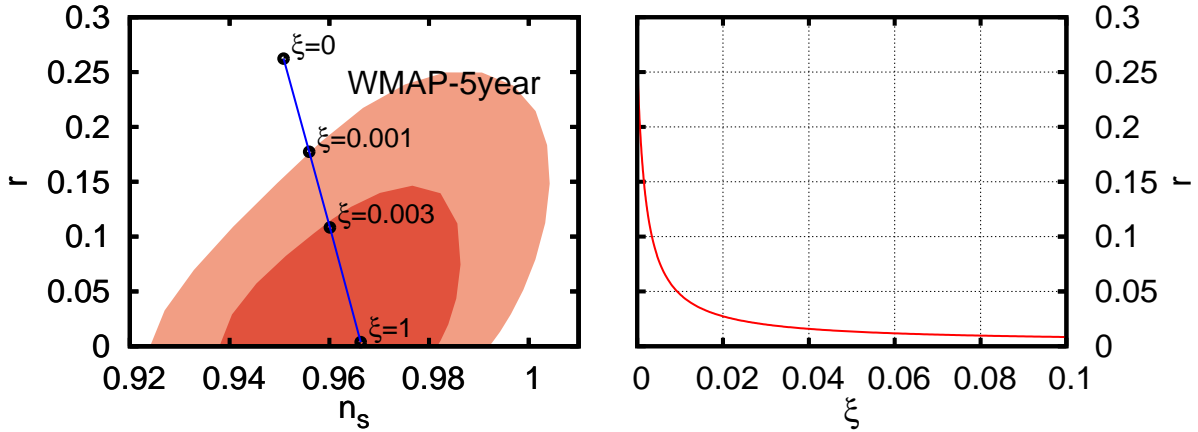}
  \caption{Left: WMAP5 preferred region for the spectral index $n_s$
    and tensor-to-scalar ratio $r$ with the predictions form the
    non-minimally coupled inflation for several $\xi$.  Note, that the
    points for $\xi>1$ are indistinguishable from the one with
    $\xi=1$.  Right: scalar-ro-tensor ratio dependance on the
    non-minimal coupling constant $\xi$.}
  \label{fig:wmap}
\end{figure}

The spectral index and tensor-to-scalar ratio are independent of the
coupling constants in the limit of large $\xi$,
$n_s\simeq1-8(4N+9)/(4N+3)^2\simeq0.97$,
$r\simeq192/(4N+3)^2\simeq0.0033$.  The predicted values are well
within one sigma of the current WMAP measurements
\cite{Komatsu:2008hk}, see Fig.~\ref{fig:wmap}.

A very interesting candidate for the inflaton in this case is just the
SM Higgs field.  Though its self interaction is of the order of unity,
sufficiently large $\xi\sim10^4$, \eqref{eq:9}, allows it to drive
inflation, and no new fields have to be added to the SM.

The interesting problem here is the radiative corrections to the
potential form the other fields of the SM, like the top quark and the
gauge bosons.  There are several ways of addressing this problem, that
differ by the frame used to define he cut off energy (or normalization
point).  In \cite{Bezrukov:2007ep,Bezrukov:2008cq} the cut off
effectively corresponds to constant Planck mass in the Einstein frame,
leading to corrections that do not spoil the flatness of the potential
(\ref{eq:U(chi)}), while \cite{Barvinsky:2008ia} advocates that
constant Plank mass cut off should be taken in the Jordan frame,
leading to larger corrections, but, interestingly, still allowing for
successful inflation, only for rather heavy Higgs mass.  The proper
way of taking these corrections into account is an important question
open for further discussion.

\section{Small non-minimal coupling}

However, large non-minimal coupling $\xi$ is not the only possibility.
As far as for zero $\xi$ the amount of tensor modes predicted is too
large, and it is very small for large $\xi$, one can ask a question
what is the minimal value of $\xi$ that suppresses the tensor mode
generation just enough to put them into the WMAP5 preferred region.
To answer this question, let us write the expressions for the spectral
index and tensor-to-scalar ratio in the limit of small non-minimal
coupling constant $\xi$.

\begin{figure}
  \centering
  \includegraphics[width=0.5\textwidth]{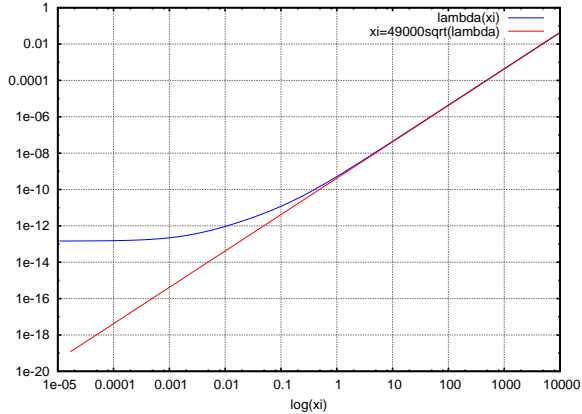}
  \caption{Dependence of the quartic self-coupling $\lambda$ on the
    non-minimal coupling parameter $\xi$, deduced from requirement of
    the correct normalization of the density perturbations.}
  \label{fig:lxi}
\end{figure}

The slow roll parameters $\epsilon$ and $\eta$ are
\cite{Kaiser:1994vs,Tsujikawa:2004my}
\begin{gather}
  \label{eps}
  \epsilon={8\Mpl^4\over\phi^2(\Mpl^2+{\xi\phi^2(1+6\xi)})}
  \;,\\
  \label{eta}
  \eta = \frac{4M_P^2(3M_P^4+\xi M_P^2\phi^2(1+12\xi)-2\xi^2\phi^4(1+6\xi))}
              {\phi^2(M_P^2+\xi\phi^2(1+6\xi))^2}
              \;.
\end{gather}
The end of the slow-roll regime ($\epsilon=1$) corresponds to the
field value $\phi_e$
\begin{equation}
  \begin{split}
    {\xi\phi^2_\mathrm{e}\over \Mpl^2}=&{1\over
      2(1+6\xi)}\left(\sqrt{192\xi^2+32\xi+1}-1\right)\\&\approx
    8\xi+{\mathcal O}(\xi^2),~(\xi\ll 1)
    \;.
  \end{split}
  \label{condition}
\end{equation}
The number of e-foldings that happened when the field changed its
value from $\phi_N$ to $\phi_e$ is
\begin{align}
  \label{ef}
  N & ={1\over\Mpl^2}\int^{\phi_N}_{\phi_\mathrm{e}}
  {V\over (dV/d\phi)}\left({d\hat\phi\over d\phi}\right)^2d\phi
  \\
  & ={1\over8}\left[\frac{\phi_N^2-\phi^2_\mathrm{e}}{\Mpl^2}(1+6\xi)
    -6\ln\left({\Mpl^2+\xi\phi_N^2} \over
      {\Mpl^2+\xi\phi^2_\mathrm{e}}\right)\right]
  \;.\notag
\end{align}
For small $\xi\ll1$ it reduces to $\phi_N\simeq\sqrt{8 (N+1)} M_P$.
Then it is easy to calculate the tensor-to scalar ratio
\cite{Komatsu:2008hk}
\begin{equation}
  \begin{split}
    r=16\epsilon&={128\Mpl^4\over\phi^2_N(\Mpl^2+{\xi\phi^2_N(1+6\xi)})}
    \\
    &\simeq {16\over (N+1)(1+8(N+1)\xi(1+6\xi))}
    \;.
  \end{split}
  \label{beta}
\end{equation}
The spectral index can be obtained as $n_s - 1 = -6\epsilon+2\eta$.
The exact formulas are not hard to get by combining \eqref{eps},
\eqref{eta}, \eqref{condition}, and \eqref{ef}, but they are slightly
cumbersome and not very instructive, though they were used to plot the
figures.  Fig.~\ref{fig:wmap} gives the results for several values of
$\xi$ together with the WMAP5 preferred region \cite{Komatsu:2008hk}.
On can see, that for $\xi>0.001$ and $\xi>0.003$ the predictions
enters the $2\sigma$ and $1\sigma$ contours, respectively.  The
quartic coupling constant $\lambda$ should still be extremely small
for these values to satisfy the COBE normalization
$U/\epsilon=(0.027M_P)^4$, see Fig. \ref{fig:lxi}.

One can even argue, that for ``natural'' value for the non-minimal
coupling constant of the order of $1$, the amount of tensor
perturbations generated is extremely small, so it is natural
\emph{not} to expect the large tensor modes form the inflation,
contrary to the usual conclusion for large single field inflation.

\section{Conclusions}

We have demonstrated that the non-minimal coupling of the scalar field
to gravity loosens the bounds on the field self-coupling constant
required for successful inflation, and reduces the amount of tensor
modes produced.  Large non-minimal coupling (of the order of $10^4$)
allows to use the SM model Higgs field for inflation.  At the same
time even very small coupling $\xi\gtrsim10^{-3}$ makes inflaton with
quartic potential compatible with the CMB observations.  For the
coupling constant $\xi\sim1$ very small amount of tensor modes is
expected for the quartic inflation.  Interestingly, that at the same
time the non-minimally coupled inflation with quadratic potential and
$\xi\sim1$ is no longer compatible with observations, generating too
small spectral index, or even leading potential with runaway behaviour
(see \cite{Tsujikawa:2004my}).

As a summary, adding non-minimal coupling constant changes the usual
expectations from inflation a lot, and lead to interesting
predictions.  One interesting application to the inflation in the
$\nu$MSM model was made in \cite{Anisimov:2008qs}.

Author is grateful to M.~Shaposhnikov for numerous helpful
discussions.


\end{document}